\documentclass{jfm}
\usepackage{graphicx}
\usepackage{epstopdf, epsfig}
\usepackage{xcolor}
\usepackage{color, soul}
\usepackage{multirow}

\shorttitle{Inhomogeneous boundary conditions}
\shortauthor{Fei Wang, Shi-Di Huang and Ke-Qing Xia}

\title{Thermal convection with mixed thermal boundary conditions: Effects of insulating lids at the top}

\author{Fei Wang\aff{1}
  ,
  Shi-Di Huang\aff{1}
 \and Ke-Qing Xia\aff{1}
 \corresp{\email{kxia@cuhk.edu.hk}}
 }

\affiliation{\aff{1}Department of Physics, The Chinese University of Hong Kong,
Shatin, Hong Kong, China
}

\begin{document}

\setstcolor{red}

\maketitle

\begin{abstract}
The effects of insulating lids on the convection beneath were investigated experimentally using rectangular convection cells in the flux Rayleigh number range $2.3\times10^{9}\leq Ra_F \leq 1.8\times10^{11}$ and cylindrical cells in the range $1.4\times10^{10}\leq Ra_F \leq 1.2\times10^{12}$ with the Prandtl number Pr fixed at 4.3. It is found that the presence of the insulating lids leads to reduction of the global heat transfer efficiency as expected, which primarily depends on the insulating area but is insensitive to the detailed insulating patterns. 
At the leading order level, the magnitude of temperature fluctuation in the bulk fluid is, again, found to be insensitive to the insulating pattern and mainly depends on the insulating area; while the temperature probability density function (PDF) in the bulk is essentially invariant with respect to both insulating area and the spatial pattern of the lids. 
The flow dynamics, on the other hand, is sensitive to both the covering area and the spatial distribution of the lids. At fixed $Ra_F$, the flow strength is found to increase with increasing insulating area so as to transfer the same amount of heat through a smaller cooling area.  Moreover, for a constant insulating area, a symmetric insulating pattern results in a symmetric flow pattern, i.e. double-roll structure; whereas asymmetric insulating pattern leads to asymmetric flow, i.e. single-roll structure. It is further found that the symmetry breaking of the insulating pattern leads to a stronger flow that enhances the horizontal velocity more than the vertical one.

\end{abstract}

\begin{keywords}
%Authors should not enter keywords on the manuscript, as these must be chosen by the author during the online submission process and will then be added during the typesetting process (see http://journals.cambridge.org/data/\linebreak[3]relatedlink/jfm-\linebreak[3]keywords.pdf for the full list)
\end{keywords}

\section{Introduction}
As a simplified model for convective flows in nature, turbulent Rayleigh-B\'{e}nard (RB) convection, a fluid layer heated from below and cooled from above, has been extensively studied in the past years (see \citet*{Ahlers2009, LohseXia2010, Chilla2012, XIA2013} for reviews). However, while thermal uniformity (i.e. either constant temperature or constant flux that are spatially uniform across the plates) has been the standard choice for the top and bottom boundary conditions in this idealised model, the thermal forcings in most natural flows are in fact inhomogeneous at the boundaries. One example is the mantle convection with a combination of continental and oceanic lithospheres acting as the top boundary. Because the mantle heat flux through the continental lithosphere ($\sim0.01\,\rm{Wm^{-2}}$) is about one order smaller than that through the oceanic lithosphere ($\sim0.1\, \rm{Wm^{-2}}$), the continental lithosphere is generally viewed as a thermally insulating lid at the top of mantle convection \citep*{Pinet1991, Guillou1994}. Another example is the oceanic circulation with sea ice floating on the surface. It has been estimated that the atmosphere-ocean heat transfer through the thick ice is only $\sim5\,\rm{Wm^{-2}}$, two orders smaller than the $600\,\rm{Wm^{-2}}$ through the open surface \citep*{Maykut1982, Maykut1986, Andreas1986}. Therefore, the sea ice also acts as a thermally insulating lid at the top of the oceanic circulation. To understand how these insulating continents and sea ices influence the behaviours of mantle flow and oceanic circulation, many efforts including theoretical modelling, numerical simulation and experiment have been carried out, which can be broadly divided into two groups.

The first group is using movable insulating lids and investigating their effects on the flow dynamics in mantle convection \citep*{Bott1964, Lowman1993, Lowman1995, Gurnis1988, ZhangJun2000, King2002, Cooper2004, ZhongJQ2005, Trubitsyn2008}. A major finding of this type of study is that because the continental plates inhibit mantle cooling, they would either aggregate over the cold downwelling regions or fragment into smaller plates owing to the convective flow beneath, and these changes in the insulating continent would in turn induce changes in the dynamics of mantle flow, which is believed to be responsible for the so-called Wilson Cycle in the earth history \citep*{Wilson1966}. The second group aims to study the insulating effects on the heat transport using fixed lids. It is generally concluded that, for a fixed insulating area, convective flow with several smaller lids at the top would have a larger heat transfer efficiency than that with one big lid of equivalent total covering area \citep*{Cooper2013, Ripesi2014}. However, it is not clear whether the flow dynamics has been changed for these cases. Moreover, as far as we are aware, no experimental work studying the heat transport in turbulent convection with insulating lids at the top has been made so far. This motivates us to carry out the present study. 

A standard turbulent Rayleigh-B\'{e}nard (RB) convection system with heat flux supplied at the bottom and removed from the top was used. By partially insulating the top boundary with fixed lids, we study the response of the heat transport and flow dynamics to the fractional insulating area and detailed insulating pattern. There are five control parameters in the experiment. The first one is the flux Rayleigh number $Ra_{F}=\alpha gFH^4/(\chi\nu\kappa)$; $Ra_{F}$ is used here since we compare cells with varying area of insulating lids under the same heat flux input from the bottom. The other control parameters are the Prandtl number $Pr=\nu/\kappa$, the lateral aspect ratio $\Gamma=L/H$, the fractional insulating area $\xi=S_{A}/S$ and the symmetry parameter $\Lambda$ of the insulating pattern. Here $F$ is the heat flux input from the bottom heating plate, $g$ is the gravitational acceleration, and $\alpha$, $\nu$, $\kappa$ and $\chi$ are respectively the thermal expansion coefficient, the kinematic viscosity, the thermal diffusivity and the thermal conductivity of the working fluid (distilled water in the present case); $S_{A}$ and $S$ are respectively the combined area of insulating lids and total top surface area; $\Lambda$ is a parameter characterising the symmetry of the insulating pattern with $\Lambda=0$ ($\Lambda=1$) standing for symmetric (asymmetric) arrangement of the insulating lids (or insulating pattern). We found that the insulating pattern, whether symmetric or not, does not influence the heat transport, in contrast to the previous numerical results \citep*{Cooper2013, Ripesi2014}. Another interesting finding of our study is that convective flows with the same heat transfer efficiency not only have different flow structures but also very different flow strengths. Flows subject to a symmetric insulating pattern ($\Lambda=0$) are weaker than those subject to an asymmetric insulating pattern ($\Lambda=1$). These results provide new understandings of thermal convection with insulating lids.

The rest of the paper is organised as follows. The experimental setups and the measurement techniques used in this study are introduced in \S\ref{sec:Setups}. The main results and discussions are presented in \S\ref{sec:Result}. We summarise and conclude in \S\ref{sec:Conclusion}.

\section{Experiments}\label{sec:Setups}

The experiments were conducted in convection cells of two different geometries. One with  rectangular geometry and the other with a cylindrical one.

\subsection{Rectangular cell}\label{sec:SetupsRec}

\begin{figure}
  \centerline{\includegraphics[width=18pc]{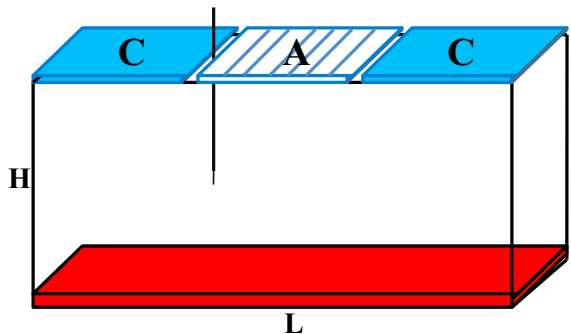}}% Images in 100% size
  \caption{A schematic drawing of the rectangular cell with a CAC configuration for the top boundary. See text for details.}
\label{fig:apparatus}
\end{figure}

A schematic drawing of the rectangular convection cell used in the first experiment is shown in Figure \ref{fig:apparatus}, which has dimensions (in cm) of $25~(L)\times12.5~(H)\times7.5~(W)$. A constant heat flux was uniformly supplied at the bottom plate, which is made of copper. The top boundary was a combination of copper plates (denoted by `C') and adiabatic plates (denoted by `A'). Specially, the `C' plate was connected to a refrigerated circulator to maintain a constant temperature, and the `A' plate was made of plexiglass with the thermal conductivity about three orders smaller than that of copper. Therefore, the top boundary covered by the `A' plate can be viewed as thermally insulating and the heat flux inputted at the bottom plate is primarily extracted from the `C' plate at the top.

To investigate the insulating effects of the `A' plates on the convection, five configurations of the top boundary, i.e. CCC, CCA, CAC, CAA and ACA, corresponding to $\xi$=0, 1/3, 1/3, 2/3 and 2/3, were studied in the parameter range $2.3\times10^{9}\leq Ra_F \leq 1.8\times10^{11}$ with the bulk temperature of the fluid  maintained at $40.0\,^\circ\hspace{-0.09em}\mathrm{C}$, corresponding to $Pr=4.3$. The lateral aspect ratio in the experiments was fixed at $\Gamma=2$.

The influence of the insulating lids (i.e. `A' plates) on the flow dynamics of the convection was investigated by carrying out 2-D velocity measurements in the mid-plane between the front and back sidewalls of the convection cell at $Ra_F=1.1\times10^{11}$, using a particle image velocimetry (PIV) system that has been introduced elsewhere \citep*{XiaPRE2003}. Briefly, the flow was seeded with particles of 50 $\rm{\mu m}$  diameter and was illuminated by a lasersheet of thickness $\sim$1 mm. Each velocity map contains $117\times57$ vectors and 7200 velocity maps were acquired with the sampling rate of $\sim$2 Hz. The response of heat transfer to the insulating lids relies on precise temperature measurement. The temperatures of all the plates (including the `A' plates) were measured by embedded thermistors with a head of 2.4 mm in diameter (Omega 44031), and the temperature of the bulk fluid was measured by a small thermistor with a head of diameter 0.38 mm and a time constant of 30 ms in water. All the thermistors were calibrated individually to ensure an accuracy better than $0.01 \,^\circ\hspace{-0.09em}\mathrm{C}$. To eliminate the influence from the environment, the convection cells were wrapped by several layers of polystyrene foam sheets and placed in a thermostat box with a temperature stability better than $0.1 \,^\circ\hspace{-0.09em}\mathrm{C}$.

\subsection{Cylindrical cell}
\begin{figure}
  \centerline{\includegraphics[width=30pc]{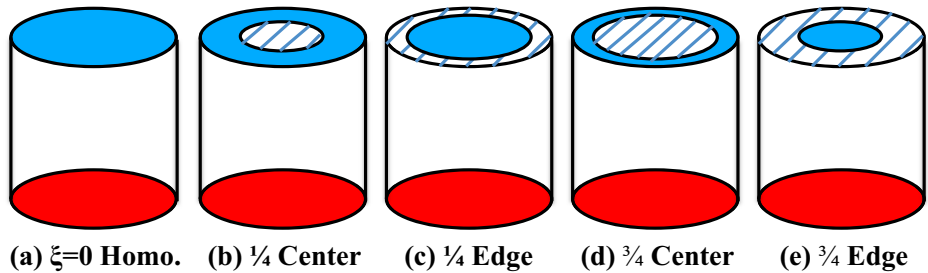}}% Images in 100% size
  \caption{Schematic drawings of (a) the convection cell with homogeneous boundary condition at the top, i.e. $\xi=0$, and  (b-e) convection cells with different insulating patterns at the top plates. Each pattern is named as ``insulating area + insulating location''. For example, ``1/4 center" means that the insulating material covers the central region with an area equals to 1/4 of the total area of the top plate.}
\label{fig:apparatusCy}
\end{figure}

Cylindrical convection cells with diameter $D$ and height $H$ both equal to 20 cm were used in the second experiment and are shown in Figure \ref{fig:apparatusCy}. Different from the thermal insulating method for the rectangular cell (using plexiglass plates), part of the top plate of the cylindrical cell was insulated by coating a material with thermal conductivity $\chi = 0.016 W/(m\cdot K)$ to either the central region or the edge of the top plate as shown in Figure \ref{fig:apparatusCy}(b-e). Similarly, there are totally five configurations of the top plate used in this experiment: ``$\xi=0$ Homo.'',  `1/4 Center', `1/4 Edge', `3/4 Center', and `3/4 Edge'.

The values of the control parameters are: $1.4\times10^{10}\leq Ra_F \leq 1.2\times10^{12}$, $\xi$=0, 1/4 and 3/4, Pr=4.3 and $\Gamma=1$. Only heat transport measurements were made in this second experiment using cylindrical cells.

\section{Results and discussion}\label{sec:Result}

\subsection{Heat transfer}\label{sec:NuRect}

\begin{figure}
  \centerline{\includegraphics[width=30pc]{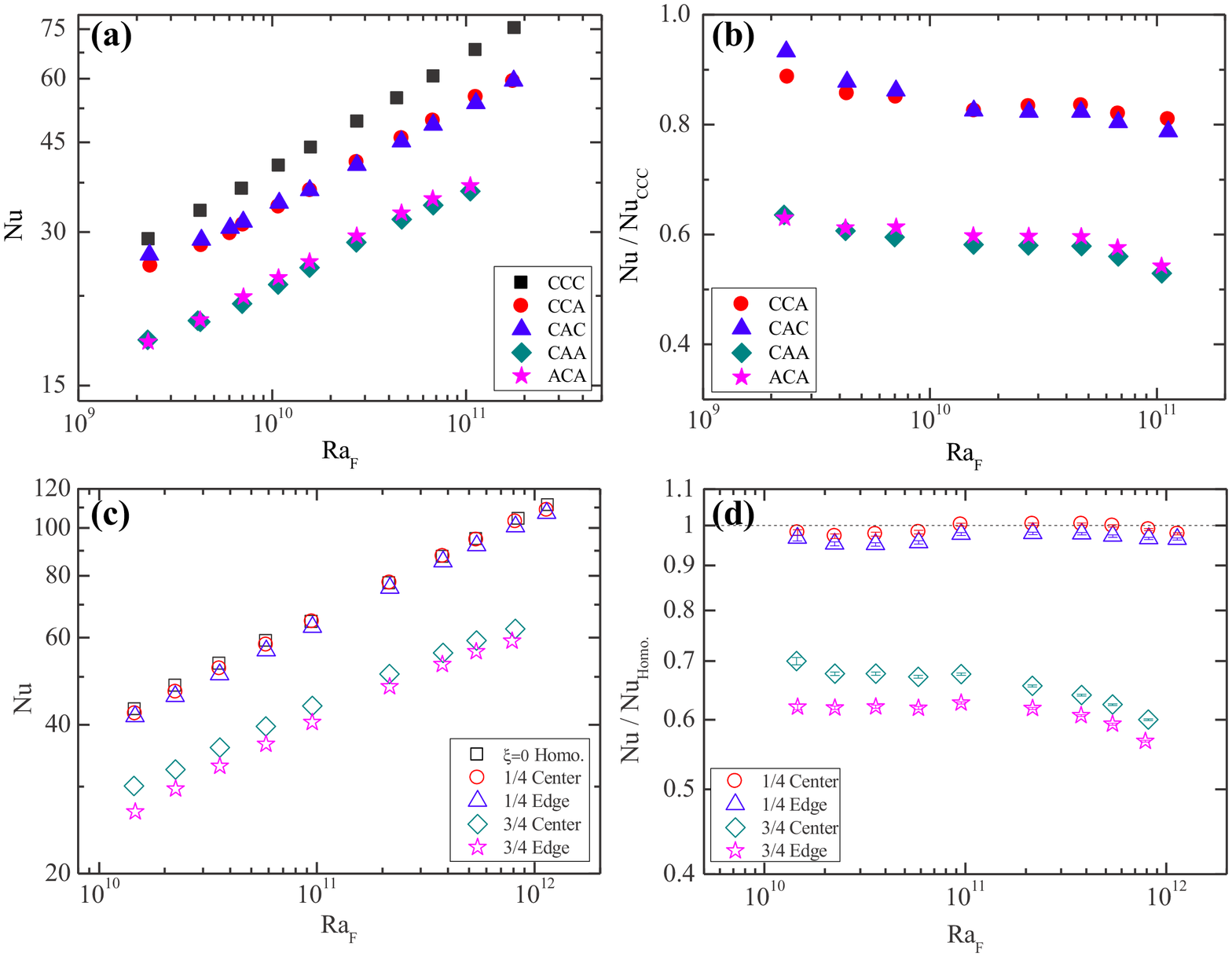}} % Images in 100% size
  \caption{(a) The global heat transfer efficiency $Nu$ as a function of the flux Rayleigh number $Ra_F$  for different configurations of the rectangular cell. (b) The corresponding data normalised by $Nu$ in the `CCC' case. The lower panel corresponds to (c) Nu and (d) normalised Nu in the cylindrical cells. The dash line with value $Nu / Nu_{Homo.}=1$ in panel (d) is plotted for reference.
  }
\label{fig:Nu}
\end{figure}

\begin{figure}
 \centerline{\includegraphics[width=32pc]{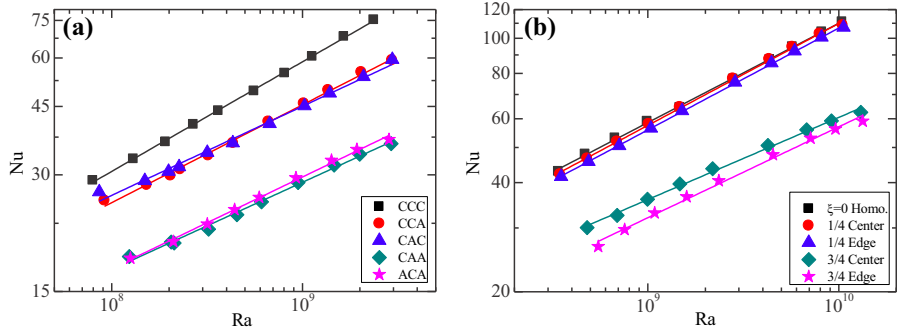}}% Images in 100% size
  \caption{ The global heat transfer efficiency $Nu$ as a function of the Rayleigh number $Ra$  for different configurations in (a) the rectangular cells and (b) the cylindrical cells. The solid lines correspond to power law fittings with the fitting parameters list in Table \ref{NuRaScaling}.}
\label{fig:NuRaScaling}
\end{figure}

We first examine the influence of insulating lids on the global heat transfer efficiency, defined by the Nusselt number $Nu=F/(\chi \Delta T/H)$, where $\Delta T$ is the temperature difference between the `C' plates and the bottom plate. As shown in Figure \ref{fig:Nu}(a-b) for Nu in the rectangular cells, the heat transfer efficiency decreases with the presence of the `A' plate and more `A' plates result in larger reduction of $Nu$ (the average reduction of Nu is about 16\% when adding one `A' plate and 40\% when adding two `A' plates.); however, the location of `A' plates seems to play a negligible role on the heat transport reduction behaviour (the relative Nu difference between `CCA' and `CAC' cases are within 3\% with the exception of the datum at the lowest $Ra_F$ which is about 5\%. The Nu difference between `CAA' and `ACA' cases is also within 3\%.). For the cylindrical cells (see Figure \ref{fig:Nu}(c-d)), similar phenomenon is observed: i.e., the leading order change of Nu is insensitive to the spatial arrangement of the insulating lids, but primarily depends on the fractional insulating area. One may find that the reductions of Nu in the cylindrical cell are not as much as those in the rectangular cell even though the fractional insulating area in the cylindrical cell are slightly larger. Possible reasons for this are that the insulating materials as well as the construction of the insulating lids used for the cylindrical and rectangular cells are different. For example, the `A' plates (Plexiglass) used in the rectangular case is much thicker than the coating of the insulating material in the cylindrical cell. To understand how the insulating lids changes the scaling behaviour of global heat transfer, we show in Figure \ref{fig:NuRaScaling} log-log plots of Nu versus Rayleigh number $Ra$($=\alpha g\Delta TH^3/(\nu\kappa)$) for rectangular and cylindrical cells respectively. The corresponding power law fitting parameters (scaling exponent and prefactor) are listed in Table \ref{NuRaScaling}. It is seen that, in general, the addition of insulating lids at the top not only decreases the magnitude of Nu but also reduces its scaling exponent with Ra. An exception is in cylindrical cells with $1/4$ of insulating area, within experimental uncertainty both the scaling exponent and the magnitude of Nu are essentially the same as the uncovered (homogeneous) case. Clearly, more precise measurements are needed to ascertain what exactly happens in this case.  

The weak dependence of heat transfer efficiency Nu on the insulating pattern contrasts strongly to the simulation work by \citet*{Cooper2013}, who find that the dependence of heat transfer on the lids configuration is strong for $\xi>50\%$. In their study, a significant enhancement of heat transport up to 44\% was found due to the breakup of a super lid ($\xi=75\%$) on the top surface into several smaller lids. Another 2D simulation work by \citet*{Ripesi2014} gives a similar conclusion. They found that for a constant $\xi=0.5$, there will be a systematical increase of the heat flux with a decrease of the lid size $\lambda$ until $\lambda$ becomes comparable with the thermal boundary layer thickness.  According to their findings, one would expect an enhancement of Nu in the `ACA' configuration compared to that in the `CAA' case, since the size of the `A' plate in the present case is much larger than the thermal boundary layer thickness. However, the Nu data in the two configurations are almost the same in our experiment. The reason for this discrepancy between the experiment and simulation is unclear and requires further study.

\begin{table}\small
\begin{center}
              \def~{\hphantom{0}}
\begin{tabular}{ccc}
\hline
\hline
\multirow{2}{*}{Configuration}   &  \multicolumn{2}{c}{$Nu=C Ra^{\alpha}$}  \\
\cline{2-3}                       
   &  $\alpha$  & $C$ \\
\hline
CCC       &  \rm{$0.28\pm0.01$}        & 0.17  \\ 
CCA        &  \rm{$0.25\pm0.03$}   & 0.24   \\           
CAC         &  \rm{$0.23\pm0.04$}     & 0.39   \\      
CAA         &  \rm{$0.22\pm0.03$}      & 0.27  \\                                 
ACA        &  \rm{$0.23\pm0.01$}      & 0.22  \\ 
\hline
$\xi=0$ Homo.   & \rm{$0.27\pm0.01$}   &  0.2 \\
1/4 Center     &  \rm{$0.28\pm0.02$}  &  0.17  \\
1/4 Edge    & \rm{$0.28\pm0.01$}   &  0.17  \\
3/4 Center   & \rm{$0.23\pm0.04$}   &  0.33  \\
3/4 Edge   &   \rm{$0.25\pm0.04$}  &  0.18   \\
\hline
\hline
\end{tabular}
\vspace{4 mm}
\caption{\label{NuRaScaling} Fitting parameters of the $Nu-Ra$ scaling in Figure \ref{fig:NuRaScaling}.}
\end{center}
\end{table}

\begin{figure}
  \centerline{\includegraphics[width=32pc]{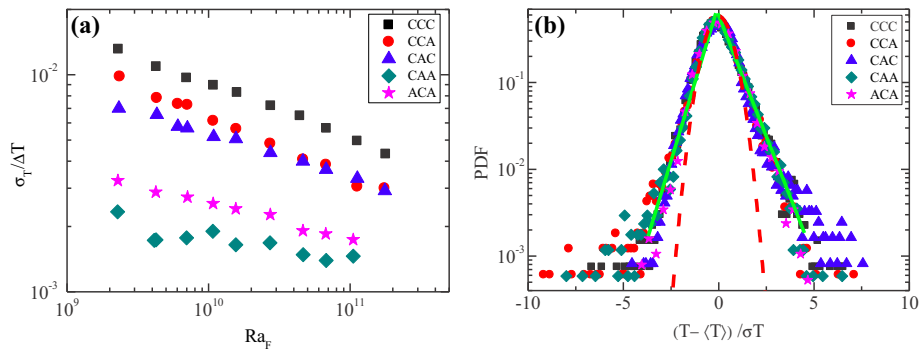}}% Images in 100% size
  \caption{(a) The bulk temperature RMS $\sigma_T$  normalised by the global temperature difference $\Delta T$ as a function of $Ra_F$. (b) The PDFs of the normalised bulk temperature fluctuations at $Ra_F=1.1\times10^{11}$ for five different configurations of the rectangular cell. The green solid and red dash lines indicate the exponential distribution and Gaussian distribution, respectively. }
\label{fig:Trms}
\end{figure}

To see how the insulating lids affect the statistical properties of the local temperature field, we examine the temperature fluctuations measured at the bulk region. It is seen in Figure \ref{fig:Trms}(a) that the magnitude of temperature fluctuations behave similarly as the Nusselt: i.e. at the leading order level, the Root-Mean-Square (RMS) temperature is insensitive to the insulating pattern and mainly depends on the insulating area. Furthermore a larger insulating area leads to a larger reduction in the RMS temperature (average reduction of about 35\% when inserting one `A' plate and 74\% when inserting two `A' plates). The temperature PDF, on the other hand, is essentially invariant with respect to both the spatial arrangement and area of the lids. This suggests that the presence of insulating lids does not strongly change the plume dynamics in the bulk of the convective flow.

\subsection{Flow dynamics}\label{sec:Dynamics}
\begin{figure}
  \centerline{\includegraphics[width=33pc]{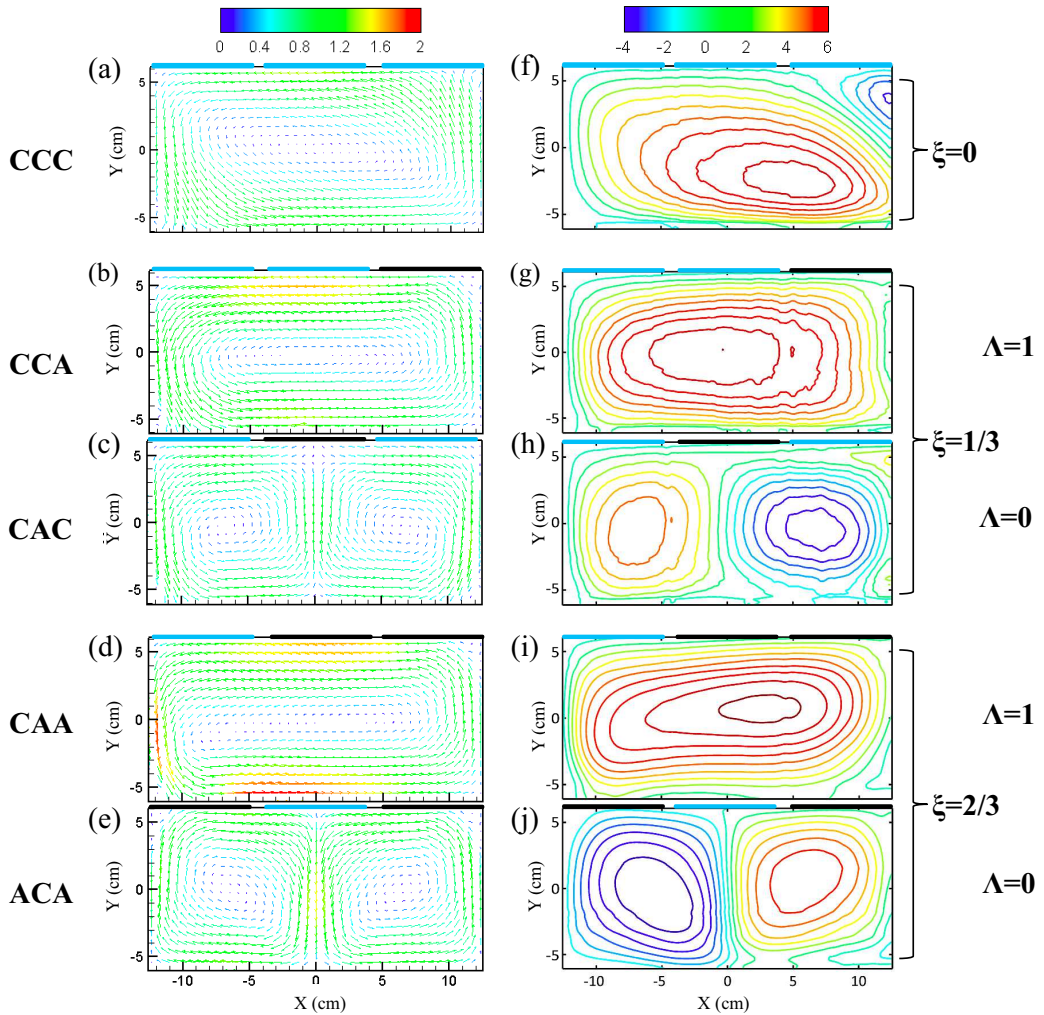}}% Images in 100% size
  \caption{(a-e): Time averaged velocity fields for five configurations at $Ra_F=1.1\times10^{11}$. The scale bar represent $(U^2+V^2)^{1/2}$ in unit of cm/s. (f-j): The corresponding contour maps of stream function with the scale bar in unit of $\rm{cm^2}/s$. The stream function values are set to zero at the cell boundaries. The contour interval is 0.6 $\rm{cm^2/s}$. Note that the `CAC' and `ACA' configurations correspond to symmetric insulating patterns ($\Lambda=0$) at the top, while the `CCA' and `CAA' configurations correspond to asymmetric insulating patterns ($\Lambda=1$).
  }
\label{fig:PIV}
\end{figure}

\begin{table}\small
\begin{center}
              \def~{\hphantom{0}}
\begin{tabular}{cccccccc}
\hline
\hline
Configuration   &  $\xi$  & $\,\,\Lambda$ \quad  &  Structure  & $V_{max}\,\rm{(cm/s)}$  &  KE $\rm{(cm^2/s^2)}$ &  $\Psi_{max}\,\rm{(cm^2/s)}$  &  $\Psi_{min}\,\rm{(cm^2/s)}$ \\
\hline
CCC       &  0        & $-$ &  SR           & 1.4                                   & 0.26                          & 5.6                    & $-$ \\
CCA        &  1/3    & 1 &  SR         & 1.7                                   & 0.37                          & 5.7                     & $-$  \\             
CAC         &  1/3     & 0  &  DR        & 1.3                                   & 0.21                          & 3.7                 & -3.1 \\         
CAA         &  2/3      & 1 &  SR        & 1.9                                   & 0.51                          & 5.9                 &  $-$ \\                                             
ACA        &  2/3      & 0 &  DR         & 1.5                                   & 0.31                          & 4.4                &  -3.9  \\
\hline
\hline
\end{tabular}
\vspace{4 mm}
\caption{\label{tab.PIVdata} Flow field quantities obtained from the PIV-measured time-averaged velocity field  for different configurations in the rectangular cell at $Ra_F =1.1\times10^{11} $. $\xi$: fractional insulating area; $\Lambda$: symmetry parameter with $\Lambda=0$ ($\Lambda=1$) stands for symmetric (asymmetric) insulating pattern; Structure: indicating whether the flow field has a single-roll (SR) or a double-roll (DR) structure; $V_{max}$: The maximum value of the velocity magnitude; KE: mean kinetic energy per unit mass $KE=(U^2+V^2)/2$; $\Psi_{max}$ ($\Psi_{min}$): the maximum (minimum) stream function with positive (negative) value stands for overturning in counter-clockwise (clockwise) direction. }
\end{center}
\end{table}

To gain a direct picture of how the convective flow has been changed due to the insulating lids, we show in Figure \ref{fig:PIV} the PIV-measured velocity fields of the rectangular cell in five configurations at $Ra_F=1.1\times10^{11}$, along with the corresponding contour maps of stream function. It is seen that the influences of the insulating lids on the flow dynamics are primarily twofold. The first one is that with the same insulating fractional area $\xi$, different arrangements of the insulating lids will result in different flow patterns. Specifically, a symmetric/asymmetric insulating pattern leads to a symmetric/asymmetric flow pattern, i.e. an asymmetric single-roll (SR)  flow structure emerges in the asymmetric CCA and CAA configurations and a symmetric double-roll (DR) flow structure emerges in the symmetric CAC and ACA cases. It is further seen from Table \ref{tab.PIVdata} that the flows subjected to symmetric insulating pattern ($\Lambda=0$) are generally less energetic than those subjected to an asymmetric insulating pattern ($\Lambda=1$), in terms of maximum velocity $V_{max}$, mean kinetic energy per unit mass KE and the maximum stream function.  The second effect is that for the same value of symmetry parameter $\Lambda$ the flow becomes stronger with increasing $\xi$. For example, for the cases with asymmetric insulating pattern ($\Lambda=1$),  $KE$,  $V_{max}$ and $\Psi_{max}$ in CCA ($\xi=1/3$) are all larger  than in  CCC ($\xi=0$); and they further increase in the CAA configuration (i.e. $\xi = 2/3$, see Table \ref{tab.PIVdata} for detailed values). Similarly, for the symmetric insulating pattern ($\Lambda=0$), these flow-strength-related quantities in the ACA configuration ($\xi = 2/3$) are larger than those in the CAC case ($\xi = 1/3$). These changes in the flow strength owing to either the area or spatial arrangement of the insulating lids are in sharp contrast to the Nu behaviour, which decreases as $\xi$ increases but are almost the same with the same $\xi$. To understand this, we note that with the presence of the insulating lids, the flow must be stronger so that it can more efficiently remove the same amount of heat flux through a smaller cooling area at the top. Therefore, at the same $Ra_F$, the flow strength is larger for larger insulating area $\xi$. As for the situation with the same $\xi$, notice that in the symmetric flow structure (DR structure) there are more fluid ``participating" the heat transport process (i.e. one more branch in the middle region of the convection cell, see Figure \ref{fig:PIV}), and also its circulating path is shorter, so the overall flow can circulate with a smaller velocity to transport the same amount of heat, compared to that in the asymmetric flow structure (SR structure). Another notable feature is that for all the cases the upwelling hot fluids are always beneath the insulating lids, whereas the downwelling cold fluids are always beneath the conducting plates, which is somewhat expected.

 \begin{figure}
  \centerline{\includegraphics[width=33pc]{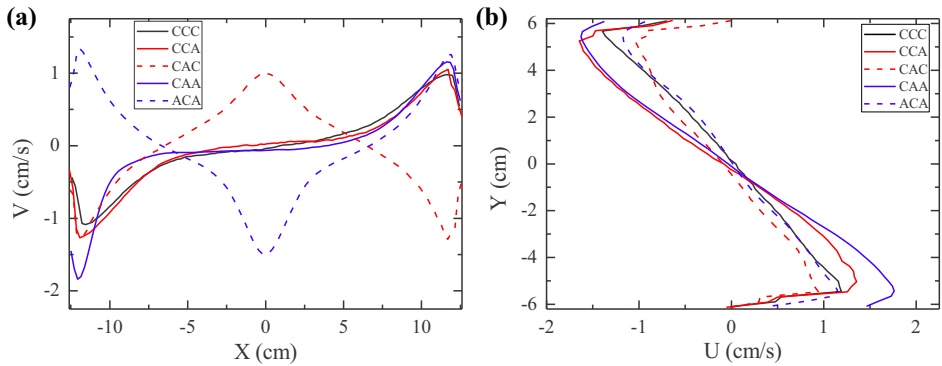}}% Images in 100% size
  \caption{(a) Horizontal cuts of the vertical velocity $V(x)$ at $Y=0$ cm. (b) Vertical cuts of the horizontal velocity $U(y)$ at $X=0$ cm for CCA and CAA cases, $X=-L/4$ for CAC case, and $X=L/4$ for the ACA case.}
\label{fig:VeloProfile}
\end{figure}

In order to further examine how the flow strength is changed in detail, we show in Figure \ref{fig:VeloProfile} the velocity profiles along the horizontal and vertical cuts. Note that the different shapes between the solid and dash lines in Figure \ref{fig:VeloProfile}(a) represent the differences between the asymmetric (single roll) and symmetric (double roll) flow structures. For the cases with different $\xi$ but the same flow structure (same value of $\Lambda$), we can see that both the vertical and horizontal velocities increase as $\xi$ increase, and the increases are much more significant in the regions away from the insulating lids. This is likely caused by a larger temperature gradient in both horizontal and vertical directions  arising from the insulating patterns. When it comes to the cases with the same $\xi$ but different symmetric parameter $\Lambda$, things are a little different. It is found that the differences in the magnitude of the horizontal velocity between the asymmetric (solid lines) and symmetric (dash lines) structures are more pronounced than those in the vertical directions. This indicates that the symmetry breaking of the insulating pattern enhances the horizontal velocity more than it does for the vertical velocity.

\section{Conclusions}\label{sec:Conclusion}

We have investigated experimentally how insulating lids influence the thermal convective flow beneath. Two sets of apparatus were used and consistent results are obtained. It is found that the leading order effect of the lids on the global heat transfer efficiency Nu depends on the fractional insulating area $\xi$ and is insensitive to the spatial arrangement of the lids. At the leading order level, the magnitude of temperature fluctuation in the bulk fluid is, again, insensitive to the insulating pattern and mainly depends on the insulating area; while the temperature PDF are essentially invariant with respect to both insulating area and the spatial pattern of the lids. However, different insulating patterns can significantly alter the flow pattern and also the flow strength, which may be related to the change in the symmetry of the thermal boundary condition by the lids. Asymmetric insulating pattern leads to stronger flow than symmetric insulating pattern does. Symmetry breaking of the insulating pattern enhances the horizontal velocity more than the vertical velocity. Moreover, in contrast to the reduction of heat transfer efficiency due to increase in the insulating area $\xi$, the flow strength can become stronger because the same amount of heat has to be removed through a smaller cooling area at the top. It is also found that the upwelling of hot fluid always occurs beneath the insulating lids, while downwelling of cold fluid always occurs beneath the conducting plate.  These findings sheds new light on the insulating effects of continental lithosphere on mantle convection (as compared to oceanic lithosphere) and have implications for how sea ice affects oceanic circulation, i.e. the breakup of sea ice may not affect the heat transport between the ocean and the atmosphere at the leading order level, but can alter the flow pattern as well as flow strength of the ocean circulation beneath.\\

We thank L. Biferale and Y.-C. Xie for useful discussions. This work was supported by the Hong Kong Research Grants Council (RGC) under grants CUHK 404513 and 14301115.

\bibliographystyle{jfm}

%\bibliography{WangFei}

\providecommand{\noopsort}[1]{}\providecommand{\singleletter}[1]{#1}%

\end{document}